\let\new=\newcommand
\new{\eq}{\begin{equation}}
\new{\en}{\end{equation}}
\new{\degree}{$^\circ$ }
 
\documentclass[12pt,preprint]{aastex}

\shorttitle{Spectral Typing of Companions}
\shortauthors{Roberts et al.}
 
\begin{document}

\title{Spectral Typing of Late Type Stellar Companions to Young Stars from Low Dispersion Near-Infrared Integral Field Unit Data}
 
\author{Lewis C. Roberts, Jr.\altaffilmark{1},  
Emily L. Rice\altaffilmark{2}, 
Charles A. Beichman\altaffilmark{1,3,4}, 
Douglas Brenner\altaffilmark{2}, 
Rick Burruss\altaffilmark{1}
Justin R. Crepp\altaffilmark{3}, 
Richard G. Dekany\altaffilmark{3}, 
Lynne A. Hillenbrand\altaffilmark{3}, 
Sasha Hinkley\altaffilmark{3}
E. Robert Ligon\altaffilmark{1},
Thomas G. Lockhart\altaffilmark{1},
David King\altaffilmark{5}, 
Stanimir Metchev\altaffilmark{6}, 
Ben R. Oppenheimer\altaffilmark{2}, 
Ian R. Parry\altaffilmark{5}, 
Laurent Pueyo\altaffilmark{7,8}, 
Jennifer E. Roberts\altaffilmark{1}, 
Michael Shao\altaffilmark{1}, 
Anand Sivaramakrishnan\altaffilmark{7},
R\'{e}mi Soummer\altaffilmark{7},
Gautam Vasisht\altaffilmark{1}, 
Fred E. Vescelus\altaffilmark{1},
J. Kent Wallace\altaffilmark{1},
Neil T. Zimmerman\altaffilmark{2,9,10} and
Chengxing Zhai\altaffilmark{1}
}

\altaffiltext{1}{Jet Propulsion Laboratory, California Institute of Technology, 4800 Oak Grove Drive, Pasadena CA 91109, USA}\email{lewis.c.roberts@jpl.nasa.gov}

\altaffiltext{2}{American Museum of Natural History, Central Park West at 79th Street, New York, NY 10024, USA}

\altaffiltext{3}{California Institute of Technology, 1200 E. California Blvd., Pasadena, CA 91125, USA}

\altaffiltext{4}{NASA Exoplanet Science Institute, 770 S. Wilson Avenue, Pasadena, CA 911225, USA}

\altaffiltext{5}{Institute of Astronomy, University of Cambridge, Madingley Rd., Cambridge, CB3 OHA, UK}

\altaffiltext{6}{Department of Physics and Astronomy, State University of New York, Stony Brook, NY 
11794-3800, USA}

\altaffiltext{7}{Space Telescope Science Institute, 3700 San Martin Drive, Baltimore, MD 21218}

\altaffiltext{8}{Johns Hopkins University, 3400 N. Charles Street, Baltimore, MD 21218, USA}

\altaffiltext{9}{Department of Astronomy, Columbia University, 550 West 120th Street, New York, NY 10027, USA}

\altaffiltext{10}{Max-Plank-Institute for Astronomy, Koenigstuhl 17, 69115 Heidelberg, Germany}

%%%%%%%%%%%%%%%%%%%%%%%%%%%%%%%%%%%%%%%%%%%%%%%%%%%%%%%%%%%%%%%

\begin{abstract}

We used the Project 1640 near-infrared coronagraph and integral field spectrograph to observe 19 young solar type stars.  Five of these stars are known binary stars and we detected the late-type secondaries and were able to measure their $JH$ spectra with a resolution of $R\sim30$. The reduced, extracted, and calibrated spectra were compared to template spectra from the IRTF spectral library. With this comparison we test the accuracy and consistency of spectral type determination with the low-resolution near-infrared spectra from P1640.  Additionally, we determine effective temperature and surface gravity of the companions  by fitting synthetic spectra calculated with the {\tt PHOENIX} model atmosphere code. We also present several new epochs of astrometry of each of the systems.  Together these data increase our knowledge and understanding of the stellar make up of these systems.  In addition to the astronomical results, the analysis presented helps validate the Project 1640 data reduction and spectral extraction processes and the utility of low-resolution, near-infrared spectra for characterizing late-type companions in multiple systems.  
 
\end{abstract}

  
%%%%%%%%%%%%%%%%%%%%%%%%%%%%%%%%%%%%%%%%%%%%%%%%%%%%%%%%%%%%%%%

\section{Introduction}

Project 1640 is a near-infrared integral field spectrograph (IFS) behind an apodized pupil Lyot coronagraph at Palomar Observatory. The instrument is fully detailed in \citet{hinkley2011}. Between 2008 March and 2010 July, we carried out the Phase I survey, coupling the Project 1640 instrument with the 241-element PALAO adaptive optics (AO) system \citep{dekany1997, troy2000} on the 5.1 m Hale Telescope at Palomar Observatory. For the Phase II survey, the instrument is mated with the high-order AO system PALM-3000 \citep{bouchez2010} on the same telescope; the Phase II survey started in 2011 October. In combination with a post-coronagraphic wavefront calibration unit and image post processing \citep{crepp2011,pueyo2011}, we expect to generate contrasts of 10$^{-7}$ at a separation of 1\arcsec.  The project's goal is the detection and characterization of exoplanets and brown dwarfs.   
 
The stars in this paper were targets of the Formation and Evolution of Planetary Systems (FEPS) Legacy Science program for the Spitzer space telescope \citep{meyer2004},  selected to be solar-type stars spanning a range of ages from $\sim$3 Myr to $\sim$3 Gyr.  Those accessible from the northern hemisphere were surveyed for multiplicity by \citet{metchev2009}. During the P1640 Phase I survey, we observed a number of  young FEPS targets, as their youth makes them promising candidates for searches for exoplanets. This paper discusses five stars ((HD 77407, HD 91782, HD 112196, HD 129333, HD 135363) that have known stellar companions detected by previous AO observations. It also briefly discusses the stars that we did not detect any companions to.

This paper furthers our understanding of these five stellar systems by adding to the known astrometry, which will eventually allow for the computation of an orbital solution. We also provide estimates of the spectral types of the companions. Coupled with the eventual mass estimates of the various spectral types, this will provide a measurement of the mass ratio and contribute to our understanding of the mass ratio distribution for solar type stars. The companion mass distribution is a key test of various stellar formation models \citep{kouwenhoven2009,reggiani2011}.

In Section \ref{observations} we describe how the observations were carried out. We describe how the data were reduced and how the astrometry and spectrum of the companion were produced in Section \ref{data_analysis}.  Section \ref{stars}, discusses the results for each systems.  The results of a model based method for determining the effective temperatures and gravity of the companions are presented in Section \ref{model}.  The observations of stars that did not detect a companion are discussed in Section \ref{null_results}. Finally, Section \ref{summary} gives our summary and conclusions.

\section{New Observations and Previous Data}\label{observations}
 
All the stars were observed during the same observing run in 2009 March. HD~129333 was observed on 2009 March 14 UT; HD~91782 and HD~135363 were observed on 2009 March 16 UT. Finally, HD~77407 and HD~112196  were observed on 2009 March 17 UT. The mean seeing for each night was visually estimated to be approximately 1\arcsec, which is slightly better than the median seeing for the observatory.  The data collection plan was the same for all stars.  First, several unocculted IFS images were acquired. These were mainly used to guide the star onto the occulting spot.  After the star is occulted, multiple images were collected, each with an exposure time of 127.5 s.  We collected 15 images of HD~77407, HD~91782 and HD~112196. For HD~129333 we collected 10 image and for HD~135363 we collected only three images.  After collection, the data were processed with the method detailed in \citet{zimmerman2011}. The processing turns each IFS image into a three-dimensional data cube, with 23 narrow-band images with central wavelengths between 1.100\micron~and 1.760\micron~at 30 nm increments.

We combine our P1640 astrometric measurements with multi-epoch Palomar/PHARO and Keck/NIRC2 astrometric data collected between 2002-2004 from the survey of \citet{metchev2006}. These data were taken with the Palomar 5m telescope using  PALAO with the Palomar High-Angular Resolution Observer (PHARO) NIR camera \citep{hayward2001} and with the Keck II facility AO system \citep{wizinowich2000} using the NIRC2  camera. The PHARO distortion and pixel scale were calibrated  to a precision of 0.15\% through a combination of astrometric mask and  binary star measurements, as described in \citet{metchev2006}. Additional information on the observations and data reduction for these data is found in \citet{metchev2009}.  

\section{Data Analysis}\label{data_analysis}
 
%During this observing run, the pupil plane apodizer used had a very thin grid which created astrometric spots approximately eight magnitudes fainter than the primary star.  These were not bright enough to be useful to measure reliably, so the astrometry of the companion was done on unocculted images. A different grid that produced brighter spots was used in our results on Alcor \citep{zimmerman2010}.

Extracting spectra from IFS instruments is non-trivial with many challenging aspects including image registration and cross talk between channels. These have been described in \citet{zimmerman2011} which discusses the current P1640 data pipeline. Pipeline processing produces an image of the object in each of 23 wavebands resulting in a data cube.  For the unocculted data cubes, we measured the binary star astrometry  on each of the individual image frames with the program FITSTARS; it uses an iterative blind-deconvolution that fits the location of delta functions and their relative intensity to the data \citep{tenBrummelaar1996, tenBrummelaar2000}.  After throwing out any image frames that failed to converge to a physical solution, the position angle and separation was computed as the weighted mean of the results from all the frames. The weights were set equal to the inverse of the RMS residual of the fit, a standard output of the FITSTARS program. The error bars for position angle and separation are the weighted standard deviation of the results.  The plate scale and the position angle offset were assumed to be constant between observing runs. See \citet{zimmerman2010} for more details on the plate scale of the Project 1640 instrument. 
 
The measured astrometry is shown in Table \ref{astrometry}.  The table lists the star's Washington Double Star Catalog number, the HD number, the date of the observation in Besselian years, the position angle, separation and the instrument that was used to collect the data.  We also include the  astrometry from the survey of \citet{metchev2009}. In that paper, only the discovery astrometry was published; we include all the astrometry from the survey. For the PHARO data, we have excluded the 0.12 deg $1\sigma$ uncertainty in the detector position angle, as that uncertainty is systematic and masks the already apparent orbital motions, especially in HD~112196 AB. The astrometry published here is slightly different than that published in \citet{metchev2009}.  In \citet{metchev2006}  one of the corners of the PHARO detector was used as the origin for solving for the principal magnification component of the astrometric distortion. For the sake of absolute astrometric calibration, and for consistency with all other parts of the astrometric distortion analysis,  the current data analysis used the center of the array as the origin of the distortion.  This results in a small improvement in the astrometry.  
 
To create a spectrum of each object, we performed aperture photometry on each of the 23 images in the data cube. This was done with the \textit{aper.pro} routine which is part of the IDL astrolib\footnote{http://idlastro.gsfc.nasa.gov} and is an adaptation of \textit{DAOphot} \citep{stetson1987}.  An aperture was drawn around each object as well as an annulus. The radius of the aperture and annulus was adjusted for each star, so it avoided the other star and avoided the occultation spot.   The background was set equal to the average intensity for each pixel in the annulus and was subtracted from each pixel in the aperture.  The  remaining power was then totaled.  The total power as function of wavelength is the corresponding spectrum. The aper.pro code also produces an measurement error bar based on photon statistics and the error in the background estimation. 
 
The extracted spectrum is a convolution of the object spectrum, the instrumental spectral response function and the absorption due to the Earth's atmosphere. To remove the instrumental response and the atmospheric absorption, we used the primary in unocculted images as a reference source. We used a spectrum corresponding to the spectral type of the primary from the Infrared Telescope Facility (IRTF) Spectral Library \citep{cushing2005, rayner2009}.  We divide the extracted spectrum of the primary by the reference spectrum; the result is the spectral response function (SRF) of the instrument and the atmospheric absorption.  The companion's measured spectrum is then divided by the SRF and yields the calibrated spectrum of the companion. Even though the secondaries are visible in the unocculted images, we only measured their spectra in the occulted images, as it had much higher signal to noise.

The error bars on the final reduced spectra are the combination of the errors of the measured science spectra and the error on the SRF. The SRF errors come from the combination of the errors of the measured calibration spectra and the errors on the template spectra. The error bars of the measured spectra are dominated by the cross talk error, where light from adjacent microlenses in the IFS overlap on the focal plane \citep{zimmerman2011}. The measurement error from the aperture photometry is also very small.  We had multiple data cubes of the occulted star, which reveal the companion.  In these cases, the read noise was reduced by averaging the spectra from multiple data cubes.   For the calibration observations, where we only had one or two observations, this was not the case and the error bars on the SRF are much higher.  This turns out to be the dominant error term in the error bars of the final spectra.  

There is one astrophysical error term that is independent of our instrument. Using an IRTF standard star in the computation of the SRF rather than the true spectra of the primary causes a small error.  There are two components of this error. One is that there may be an error in the reported spectral type of the primary; it may be either later or earlier than what we use.  On most stars, the difference between various determinations of the spectral class is only a few subclasses.  In these cases there is no change in the result. A larger error in the spectral type of the primary would result in a corresponding error in the secondary's spectral type. 

\begin{figure}[htb]
  \begin{center}
\includegraphics[height=7cm]{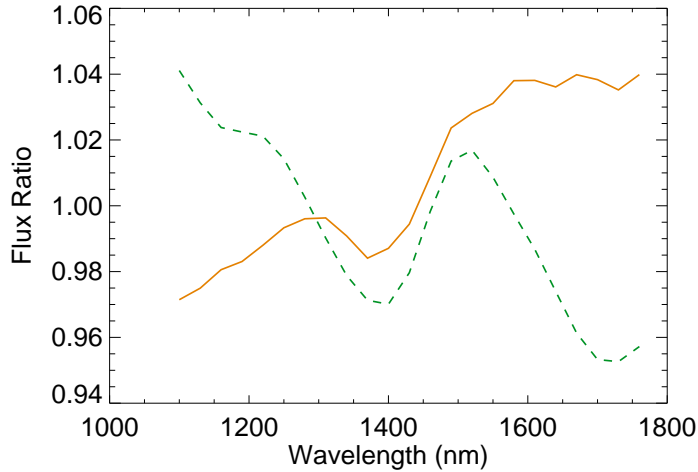}
   \end{center}
  \caption[]{ \label{template_error} The solid line (orange in online edition) is the ratio of spectra for two F8V stars, and the dashed line (green in online edition) is the ratio of two G8V stars. All four spectra were taken from the IRTF Spectral Library. The ratio of the spectra is the relative error that would be caused by using one star's spectrum to calibrate the SRF of the other.  All spectra were binned and smoothed to appear as they would if taken by P1640. 

(A color version of this figure is available in the online journal.)
}
\end{figure}

The other error component is that the template star will not have the exact same age or metallicity as the science target, and will have a slightly different spectrum. To estimate the size of this error, we found two cases, where there were multiple observations of the same spectra type in the IRTF templates.  We computed the ratio of the spectra from the same spectral type and plotted the result in Figure \ref{template_error}.  The error was $\pm$4\%, which is smaller than the cross talk error.  Also the exact shape of the template error is unknown and will vary depending on close the parameters of the science target match the template star.  For these reasons, it was left out of the formal error.  Both components of the error can be eliminated if the spectrum of the primary is known from observations with other instruments.   

Our earlier papers \citet{hinkley2010} and \citet{zimmerman2010} did not calibrate the SRF on the target star, but instead used a separate observation of a calibration star to calibrate the SRF.  As such, they had large errors where telluric water bands occur, between 1.37-1.43 \micron,~and this portion of the spectrum was not used.  Our  hope was that the self calibration process using the method described above would improve the removal of telluric lines and allow for more of the spectrum to be used.  This has proven to be moderately successful for three of the stars. For these stars, the spectra have increased error bars in the water ban. As stated above, the error bars are dominated by the SRF and the SRF is dominated by the signal to noise which is low in the water band.  

For two of the stars, self calibration did not work as well as the others. For HD~135363, the separation between the primary and secondary was small enough that there was spectral contamination, which affects calibration. In this case, the SRF had to be taken from another star. In HD~112196 the reason for the poor correction the spectrum of HD~112196 is unknown.  All of the recorded spectra show a residual spectral absorption feature in the 1.37-1.43 \micron~water band, although only three minutes separate the unocculted and the first occulted image. So it is probably not a case where the humidity changed during the observations. In the end, we did not use the portion of the HD~112196 spectra in the water band.

After the spectrum of the companion was extracted, it was then compared against the spectra in the IRTF Spectral Library, which were used as templates. Since the primaries are all young stars, we compared the spectrum against the FGKM main sequence stars and  LT brown dwarfs. The template spectra were binned and smoothed in order to produce the equivalent spectra to having the star observed by P1640 \citep{zimmerman2011}.  The spectral standards were then compared against the measured spectrum. The metric for comparison was the weighted sum of the squares of the residual (SSE). 

\begin{equation}\label{fit}
SSE=\sum_\lambda w_\lambda (S_\lambda - R_\lambda)^2,
\label{sse}
\end{equation}

\noindent where,  $S_\lambda$ is the measured spectrum at a given wavelength, $R_\lambda$ is the binned reference spectrum at the same wavelength and $w_\lambda$ is the weight at the wavelength. The best fit reference spectrum was the one with the minimum value to the metric.    The weights were set equal to the inverse of the computed error at each wavelength point. Usually several adjacent spectral types fit equally well within the error bars and produce a range of spectral types.

%%%%%%%%%%%%%%%%%%%%%%%%%%%%%%%%%%%%%%%%%%%%%%%%%%%%%%%%%%%%%%%

\section{Results}\label{stars}

\subsection{HD~77407}

HD~77407 (WDS 09035+3750) is a binary system of a young GOV and a late type companion. It is a nearby star, with a distance of 30.1~pc \citep{vanleeuwen2007}. The companion was first detected by \citet{mugrauer2004} via near-IR AO. They also measured its radial velocity over two years and show the star exhibits a long term radial velocity trend. The radial velocities have a great deal of noise, presumably from chromospheric activity of the young star.  Due to this noise, they are able to place only the broadest constraints on the mass of the companion of 0.3--0.5 M$_\sun$.  They estimate the companion spectral type to be M0V-M3V based on $H$-band photometry and evolutionary models of late-type stars. \citet{metchev2009} confirmed that the objects share common proper motion. 
 
\begin{figure}[htb]
  \begin{center}
\includegraphics[height=7cm]{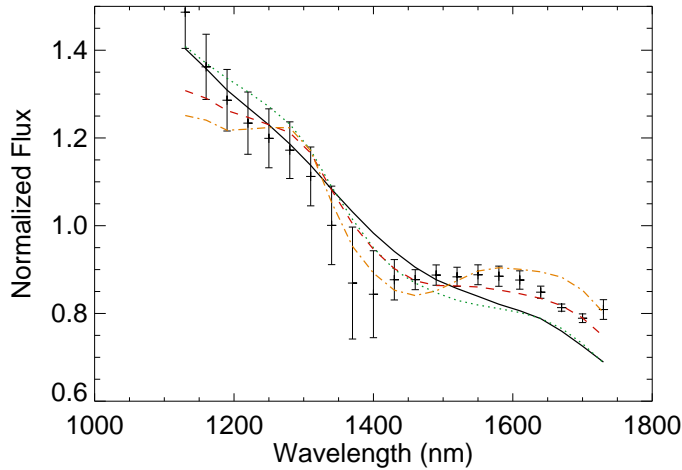}
   \end{center}
  \caption[]{ \label{hd77407_spectra}  The data points are the measured spectrum of HD~77407 B.  The black solid line  is the spectrum for an M2V star, the dotted line (green in online version) is for an M4.0V star, the dashed line (red in online version) is of an M6.0V star and the dot-dashed line (orange in online version) is for a M8V star. 

(A color version of this figure is available in the online journal.)
}
\end{figure}

The extracted spectrum of HD~77407 B is shown in Figure \ref{hd77407_spectra}; overplotted are the spectra for M2V, M4V, M6V and M8V.  These show how the spectra vary over over the range of spectral types.   Unfortunately the companion landed near the edge of the P1640 field of view.  A portion of the third Airy ring is cut off in the longer wavelength images in the data cube. This adds additional uncertainty to the extracted spectra. The best fit (via Equation \ref{fit}) for the companion's spectral type is M3V-M6V. This is later than the M0V-M3V derived by \citet{mugrauer2004}, but that determination was done  with only$H$-band photometry.  Still our results on HD~77407 B are the most uncertain of any of the stars in this paper.  Additional observations with either higher resolution or at a longer wavelength would provide more information.  

The astrometric measurements from P1640 and from \citet{metchev2006} are shown in Table \ref{astrometry}. They are consistent with the prior measurements from  \citet{mugrauer2004} and \citet{lafreniere2007}.  Using stellar masses corresponding to the spectral types of the primary and the secondary, and assuming a circular face-on orbit we estimated the orbital period of $\approx$300 years. This is an upper bound to the orbit and depending on the eccentricity of the orbit, could be significantly shorter.

%%%%%%%%%%%%%%%%%%%%%%%%%%%%%%%%%%%%%%%%%%%%%%%%%%%%%%%%%%%%%%%

\subsection{HD~91782}

HD~91782 (WDS 10368+4743) appears to be a nondescript G0 star with few published details.  Its distance is 61.4 pc \citep{vanleeuwen2007}.  The FEPS project estimated the age to be 97 Myr by an average of three techniques: rotation rate, activity, and lithium abundance. A companion was detected by \citet{metchev2009} and shown to have common proper motion. 

In the P1640 data, the companion to HD~91782 is much fainter than the primary, and it is difficult to detect the companion in many frames, especially around the water absorption bands.  Only 13 images produced meaningful results and they have a large scatter in the measured astrometry resulting in  the large error bars in Table \ref{astrometry}. The  astrometry from \citet{metchev2006} are also given in Table \ref{astrometry}. 

\begin{figure}[htb]
  \begin{center}
      \includegraphics[height=7cm]{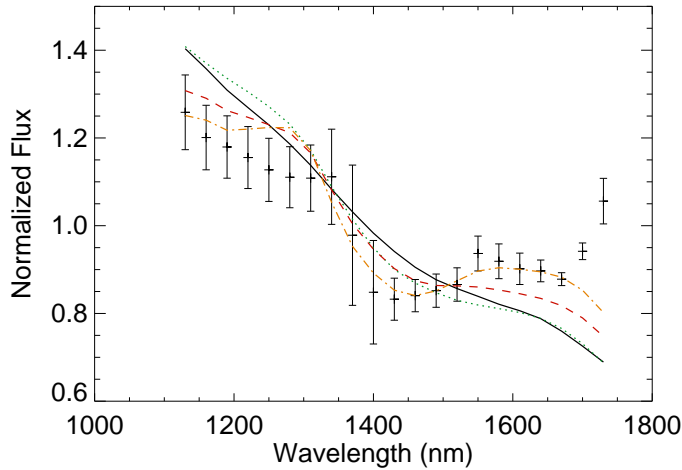}
  \end{center}
  \caption[]{ \label{hd91782_spectra} The data points are the measured spectrum of HD~91782 B.  The black solid line  is the spectrum for an M2V star, the dotted line (green in online version) is for an M4.0V star, the dashed line (red in online version) is of an M6.0V star and the dot-dashed line (orange in online version) is for a M8V star. 

(A color version of this figure is available in the online journal.)
}
\end{figure}

There is no published luminosity class for HD~91782 \citep{kharchenko2001, white2007}. Based on the age estimate, we used a main sequence G0 star from the IRTF Spectral Library as the calibration spectrum.  The extracted spectrum for the secondary is shown in Figure \ref{hd91782_spectra}.  The best fit for the spectral type of the companion is M9V.  

The astrometry for HD~91782 is shown in Table \ref{astrometry}. Only a small fraction of the orbit has been observed.    In order to get an upper bound to the orbital period, we assume it has a circular orbit with a inclination of zero. Using Kepler's third law, we get a period of $\approx$450 years.

%%%%%%%%%%%%%%%%%%%%%%%%%%%%%%%%%%%%%%%%%%%%%%%%%%%%%%%%%%%%%%%

\subsection{HD~112196}

HD~112196 (WDS 12547+2206) is a F8V  star \citep{harlan1970} at a distance of 34 pc \citep{vanleeuwen2007}. The FEPS project estimated the age to be 134 Myr (an average of calcium HK and lithium abundance methods), though the rotation-speed age measurement produced an anomalous age of 2.5 Gyr. Using isochrone fitting, \citet{holmberg2009} estimated the age to be 4.9 Gyr, with a 1$\sigma$ limit of 2.4 to 7.1 Gyr.  

A companion was detected by \citet{metchev2009}.  Using observations at multiple epochs they concluded it was a physical companion based on common proper motion. The star was observed by speckle interferometry in 1999 \citep{mason2001} and 2001 (Mason et al. in preparation), but the companion was not detected due to the faintness of the companion.

There was a single unocculted data cube of HD~112196 where both components can be detected.  The companion is easily identified in each of the data cube slice. We measured the astrometry of in each channel of the data cube with FITSTARS as described in Section \ref{data_analysis}. The data channels at wavelengths of 1100  nm and 1130  nm failed to converge. These two channels suffered from poor signal to noise due to the atmospheric water band absorption. The resulting astrometry, as well as the  astrometry from the observations in \citet{metchev2006} are shown in Table \ref{astrometry}.  

\begin{figure}[htb]
  \begin{center}
      \includegraphics[height=7cm]{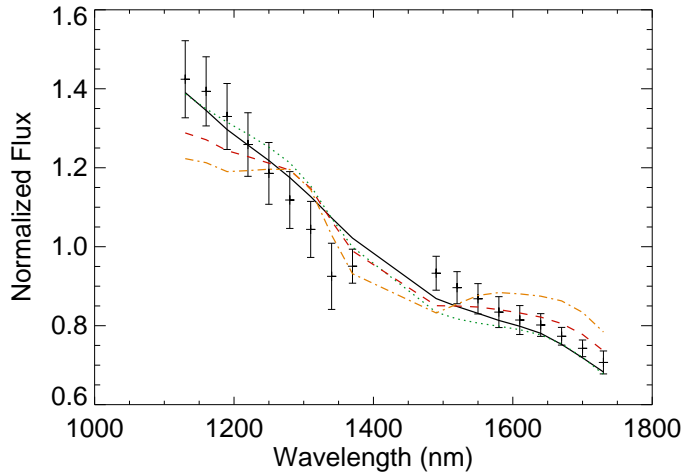}
  \end{center}
  \caption[]{ \label{hd112196_spectra} The data points are the measured spectrum of HD~112196 B.  The black solid line  is the spectrum for an M2V star, the dotted line (green in online version) is for an M4.0V star, the dashed line (red in online version) is of an M6.0V star and the dot-dashed line (orange in online version) is for a M8V star. 

(A color version of this figure is available in the online journal.)
}
\end{figure}

Unlike the other stars in this paper, the extracted spectrum for HD~112196 B shows signs of telluric water absorption after the SRF calibration process. The reason for this is unknown.  We examined the spectrum from each of the individual data cubes and they were consistent, which seems to eliminate the possibility that the water column density changed during the observations.   We did not use those data points in our spectra fitting process.  The measured spectrum for HD~112196 B is in Figure \ref{hd112196_spectra}.  The best fit for the spectral type of the companion is M2V-M3V.

The astrometry for HD~112196 is shown in Table \ref{astrometry}. The upper bound estimate for the orbital period yields a period of $\approx$320 years.

%%%%%%%%%%%%%%%%%%%%%%%%%%%%%%%%%%%%%%%%%%%%%%%%%%%%%%%%%%%%%%%

\subsection{HD~129333}

HD~129333 (WDS 14390+6417, EK Dra) is a young solar analogue \citep{dorren1994} that is highly variable and is possibly a member of the Pleiades moving group \citep{soderblom1987}. \citet{montes2001} gives a spectral type of G1.5V. \citet{duquennoy1991a} were the first to determine that this was a binary system via radial velocity measurements, and shortly thereafter \citet{duquennoy1991b} published a spectroscopic orbit of the system. \citet{metchev2004} first reported a direct image of the companion and confirmed its physical association from AO observations with the Palomar Hale telescope. Independently, \citet{konig2005} reported multi-epoch speckle interferometry observations of the companion, which they combined with radial velocity measurements to compute an orbital solution for the binary.  They estimate the period to be 45$\pm$ 5 years with an eccentricity of 0.82$\pm$0.03.  They derive a 0.5$\pm$0.1 M$_\sun$ mass for the secondary. Near-IR (J and K band) spectroscopy of HD~129333 B published in \citet{metchev2004}  sets the companion spectral type at M2$\pm$1V.
 
\begin{figure}[htb]
  \begin{center}
      \includegraphics[height=7cm]{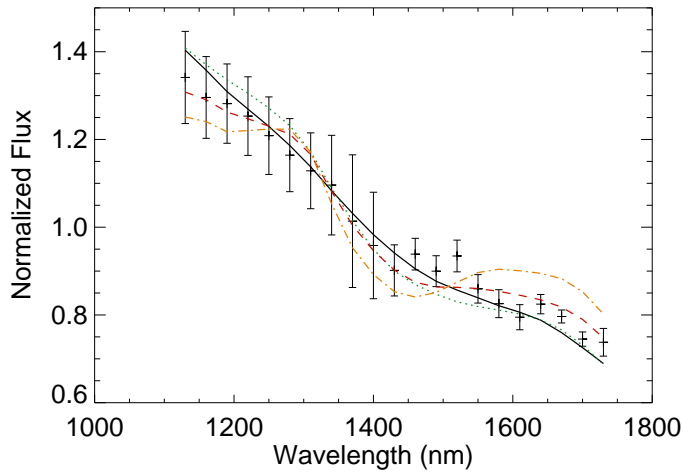}
  \end{center}
  \caption[]{ \label{hd129333_spectra} The data points are our measured spectrum of HD~129333 B. The black solid line  is the spectrum for an M2V star, the dotted line (green in online version) is for an M4.0V star, the dashed line (red in online version) is of an M6.0V star and the dot-dashed line (orange in online version) is for a M8V star. 

(A color version of this figure is available in the online journal.)
 }
\end{figure}

The extracted P1640 spectrum of the companion is shown in Figure \ref{hd129333_spectra}. The best fit for the spectral type of the companion is in the range M2V-M3.5V. This is in agreement with the M2V$\pm$1 spectral determination of \citet{metchev2004}, who produced their result using much higher resolution spectrum (R=1000).

The measured astrometry is shown in Table \ref{astrometry}.  The orbital residuals between our data points and the orbit of \citet{konig2005} are shown in Table \ref{residuals}. The 2009 observation has the largest difference in position angle, but the smallest difference in separation. It is hard to say what fraction of the difference is due to errors in the computed orbit and which is due to errors in the P1640 astrometry.   We will continue to monitor the astrometric measurements from P1640 to determine if there are any systematic errors.

%%%%%%%%%%%%%%%%%%%%%%%%%%%%%%%%%%%%%%%%%%%%%%%%%%%%%%%%%%%%%%%

\subsection{HD~135363}

HD~135363 (WDS 15079+7612) is a nearby young G5V star \citep{montes2001}.  Its distance is 29 pc \citep{vanleeuwen2007}. \citet{bubar2007} give an age of 36 $^{+14}_{-6}$ Myr   and a mass of 0.78$\pm0.01$ $M_\sun$ based on its location on the HR diagram. They determine that it is a post T-Tauri star. It is a member of the IC 2391 supercluster \citep{montes2001}. 

The system has multiple components, of which the physical nature is still uncertain for two of the pairs. \citet{metchev2009} resolved the Aa,Ab pair and showed that the two components shared common proper motion.  \citet{lepine2007} identified a common proper motion companion with a separation of  116\farcs5, the B component. The C component with a separation of 7\farcs5,  was detected by \citet{lafreniere2007}. \citet{lowrance2005} observed the star with NICMOS in 1998 and failed to resolve the Aa,Ab pair. The NICMOS coronagraph has an inner working angle of  0\farcs4, and the companion was most likely within this angle.  They detected another possible companion with a separation of 17\farcs11, the D component.  The C and D components have not been confirmed and the question remains if they are gravitationally bound to the primary. The star was observed by speckle interferometry in 2001 (Mason et al. in preparation), but no companions were detected due to either the contrast exceeding the dynamic range of the instrument or to the system having a separation wider than their speckle camera's 3\arcsec~field of view. 

\begin{figure}[htb]
  \begin{center}
      \includegraphics[height=7cm]{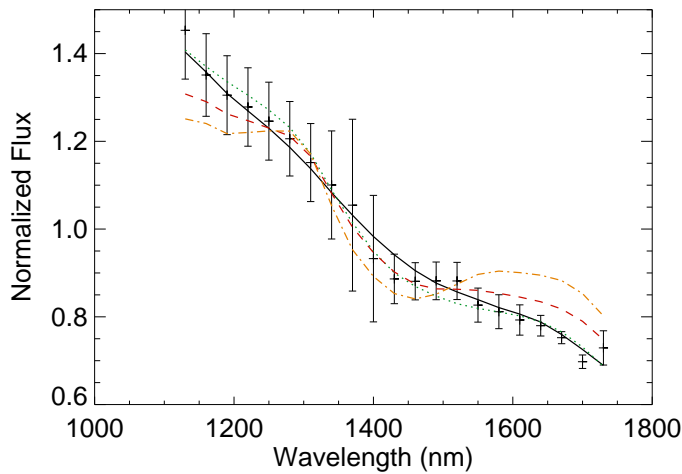}
  \end{center}
  \caption[]{ \label{hd135363_spectra} The data points are our measured spectrum of HD~135363 Ab.  The black solid line  is the spectrum for an M2V star, the dotted line (green in online version) is for an M4.0V star, the dashed line (red in online version) is of an M6.0V star and the dot-dashed line (orange in online version) is for a M8V star. 

(A color version of this figure is available in the online journal.)
 }
\end{figure}

The SRF extracted from the HD~135363 images produced unphysical results. This is probably because of contamination from the secondary.  Instead the SRF from HD~91782 taken the same night was used.  This produced the spectrum of the secondary shown in  Figure \ref{hd135363_spectra}.  The best fit for the spectral type of the companion is in the range of M2V-M4V.  

The measured astrometry is shown in Table \ref{astrometry}. From Hipparcos astrometry, \citet{makarov2005} derive a binary star period of 39.8 yrs, presumably this is from the Aa,Ab pair.  Our measurements show the companion only moved 12\degree in seven years.  An estimate of a circular face-on orbit gives a period of 29 yr. If it was such an orbit, we would have expected to see 24\% of the orbit or 87$^\circ$~in the seven years of observations.  Of the stars in this paper, HD~135363 is the highest priority for additional astrometry.  The period is probably on the order of 40 years and seven years of that has already been observed.  Observations every few years over the next few decades will enable an orbital computation.

\section{Model Comparisons}\label{model}

In addition to the spectral type analysis described above, we attempt to infer the physical properties of the companions from best fit synthetic spectra calculated with the {\tt PHOENIX} model atmosphere code  \citep{hauschildt1999}. The atmosphere structures (temperature and pressure as a function of optical depth) were calculated as described in \citet{rice2010} for a larger range of effective temperatures, in this case 1400~to~4500~K in 50~K increments. The range of surface gravities is 3.0 to 6.0 dex in cgs units), calculated in 0.1 dex increments. The {\tt PHOENIX} code produces spectra with 4~\AA~ resolution at near-infrared wavelengths, which is more than adequate for the low-resolution spectra ($\delta\lambda\sim$300~\AA) obtained with Project~1640. These data constitute the pre-calculated grid of synthetic spectra referenced below.

\subsection{Fitting Procedure}

The fitting procedure we used is based on the methods described by \citet{rice2010} for high-resolution spectra of very low mass stars and brown dwarfs. In addition to the larger range of effective temperatures considered for this study, we have improved the sophistication and efficiency of the procedure and incorporated corrections for known systematic errors introduced by the Project 1640 spectral extraction procedure (described in \citealt{zimmerman2011}). A more detailed description of the fitting procedure and the results of its application to spectral templates from late-type objects in the IRTF spectral library will be presented in Rice et al. (2012, in prep.).

 The best-fit parameters and their uncertainties are obtained in three main steps. First, a goodness of fit parameter similar to that defined in \citet{cushing2008} is determined for every pre-calculated synthetic spectrum in the model grid, nearly 2,000 spectra in total. The goodness of fit values are used to constrain the range of temperatures and gravities for the least-squared minimization calculations in the second part of the fitting procedure. The range of the constrained parameters varies for each object depending on the topology of the goodness-of-fit surface.

 The second part of the code introduces linear interpolation in flux between pre-calculated models, resulting in higher-precision best-fit parameter results than would be possible with only the pre-calculated model grid. The initial parameters of the least-squares minimizations are randomly assigned from the constrained range of temperatures and gravities. The least-squares minimization is accomplished in 10$^{5}$ iterations of the MPFIT code \citep{markwardt2009}, resulting in a distribution of best fit parameters and corresponding $\chi$$^2$-values. The parameters with the lowest $\chi$$^2$ are adopted as the starting point for the third step.

Finally, a Markov Chain Monte Carlo (MCMC) is conducted for the entire range of parameters in the model grid, and again the code linearly interpolates in flux in order to analyze an effectively continuous grid of models. The MCMC chain for each observed spectrum is 10$^{6}$ links, with the first 10\% considered ``burn-in" and not included in the final analysis. The ``jump size" (width of the normal distribution from which the new parameter was randomly selected) was optimized to produce an acceptance rate (number of accepted jumps divided by the total number of links) of 0.3--0.4 and varied from 100 to 750~K and 0.2 to 1.5~dex in surface gravity. The best-fit parameters are the modes of the 9$\times$10$^{5}$ post-burn links, which are calculated from the two-dimensional density distribution. The uncertainties are the 1-$\sigma$ width of the density distribution in each parameter. %The MCMC procedure provides the most reliable estimate of uncertainties in the best-fit parameters and provides a glimpse into possible systematic errors introduced by incomplete models. The previous fitting steps also provide a sense of the topology of the model fits, i.e., how sensitive each of the P1640 spectra are to the model parameters.

\subsection{Fitting Results}

Table \ref{results} shows our estimates for $T_{efff}$, $\log(g)$, and uncertainties inferred from the model fitting as well as the spectral types we determined from IRTF spectra. The results support potential success of efficiently and accurately inferring the physical properties of late-type companions directly from low-resolution spectra. However, the results also reveal systematic errors that need to be examined before the procedure can be reliably applied to completely unknown companions.

One object with excellent data, HD~129333 B, resulted in a best-fit temperature that is well-determined and consistent with the object's spectral type and age. The best-fit surface gravity is at the high end of the allowed range, and is unphysical according to evolutionary models. The results for this object are discussed further in Section \ref{case_study}.

For the other objects, best-fit temperatures tend to be higher than those inferred from spectral type-effective temperature conversions of late-type dwarf objects (e.g., \citep{luhman1999, luhman2003, gray2009}), although the model fitting temperature uncertainties are as high as $\pm$320~K. The best-fit temperatures of three objects (HD~77407~B, 129333~B, and HD~135363~Ab) are consistent with the spectral type determined from the P1640 data. HD~135363~Ab produced best-fit surface gravities of 5.90~dex, at the upper boundary of the models, and a slightly higher temperature than implied by its spectral type, although within the large uncertainty. HD~112196~B has the largest uncertainty in temperature of any object, and the MCMC distribution extends to $\sim$3500~K, consistent with the earliest spectral type estimate for the object. 

The latest-type object, HD~91782 B, resulted in a temperature much too high for its estimated spectral type of M9 according to all temperature scales and an unexpectedly low value for surface gravity, 3.02~dex, at the lower boundary of the surface gravity range of the models. This low surface gravity would imply an $\lesssim$1~Myr age, which is highly unlikely. The low gravity could also imply a background giant, but the star has been shown to have common proper motion by \citet{metchev2009}. It should be noted that for the latest type objects in particular, matching the near-infrared spectrum to a template spectrum is not a determination of spectral type in the strictest sense and that different wavelength regimes and even spectral resolutions could imply different physical parameters, as is the case for numerous late-M and L dwarfs. In particular, the M8.5 $\gamma$ dwarf 2MASS~J06085283$-$2753583, was described by \citet{rice2010} as 2200~K and log(g)=3.20 dex [cgs] with a moderate resolution ($R\sim$2,000) $J$-band spectrum, but high-resolution spectra ($R\sim$20,000) produced 2529~K and log(g)=3.98 dex [cgs], much more consistent with the age implied by confirmed membership in the $\beta$~Pictoris Moving Group and similarities to the well characterized TW~Hydrae member 2MASS~J12073347$-$3932540 \citep{rice2010b}.

\subsection{Case study: HD~129333 B}\label{case_study}

\begin{figure}[htb]

  \begin{center}

      \includegraphics[height=10cm]{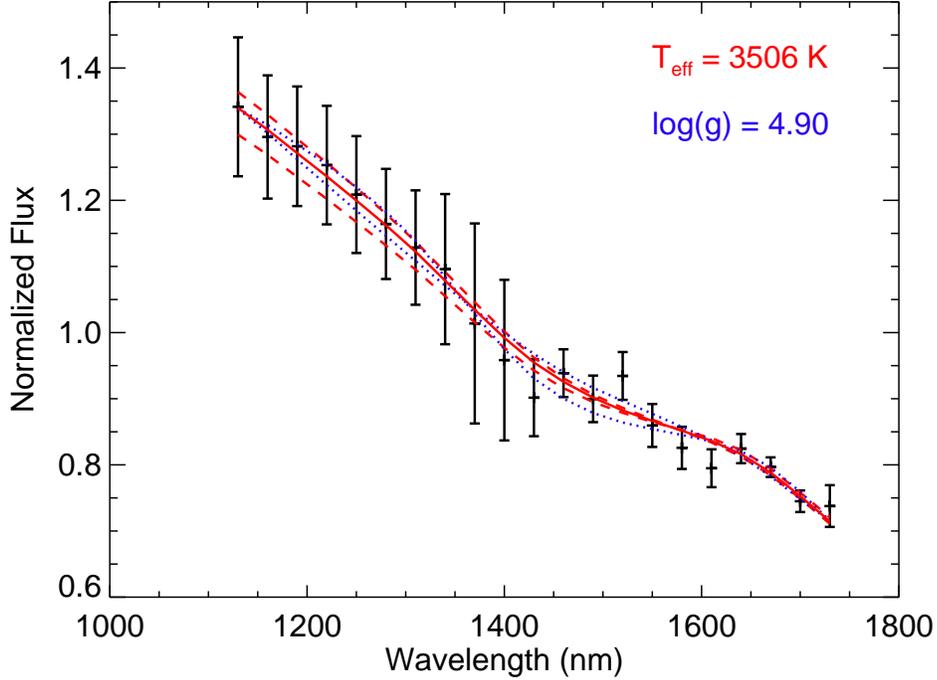}

  \end{center}

  \caption[]{ \label{hd129333b_fit} Observed P1640 spectrum of HD~129333 B (black error bars) and synthetic {\tt PHOENIX} spectrum (solid red) with best-fit parameters as determined by the spectral fitting procedure. The dashed lines (red in online version) shows two model spectra with $T_{eff}$=3506~K and surface gravities of 4.44 and 5.36 dex [cgs]. The dotted lines (blue in online version) shows two model spectra with surface gravities of 4.90 dex [cgs] and $T_{eff}$ of 3295 and 3717~K. The broken lines essentially demonstrate the 1-$\sigma$ uncertainties listed in Table~\ref{results}.}

\end{figure}

Figure~\ref{hd129333b_fit} shows the best-fit model spectrum plotted with the observed P1640 spectrum of HD~129333~B. The best fit parameters for HD~129333 B are $T_{efff}$=3506$\pm$211~K and $\log(g)$=4.90$\pm$0.46dex~[cgs]. The temperature is consistent with a M2.5$\pm$1V object according to spectral type-effective temperature conversions for either dwarf and intermediate gravity objects \citep{luhman1999}, along with the surface gravity matches the evolutionary tracks for a $\sim$100~Myr, $\sim$0.4~M$_\sun$ object. These best fit parameters are plotted at the solid red line in Figure~\ref{hd129333b_fit}). Also plotted are the 1-$\sigma$ surface gravity range (red dashed lines) at the best bit effective temperature and the 1-$\sigma$ effective temperature range at the best fit surface gravity (blue dotted lines). All spectra are normalized to the observations over the entire wavelength range. At these relatively high temperatures, the shape of  the low resolution near-infrared spectrum changes  little even with $\pm$200~K changes in temperature, but surprisingly the $J$-band slope seems somewhat sensitive to surface gravity, although all of the plotted models essentially fall within the uncertainties of the observed spectrum.

\begin{figure}[htb]

  \begin{center}

      \includegraphics[height=7cm]{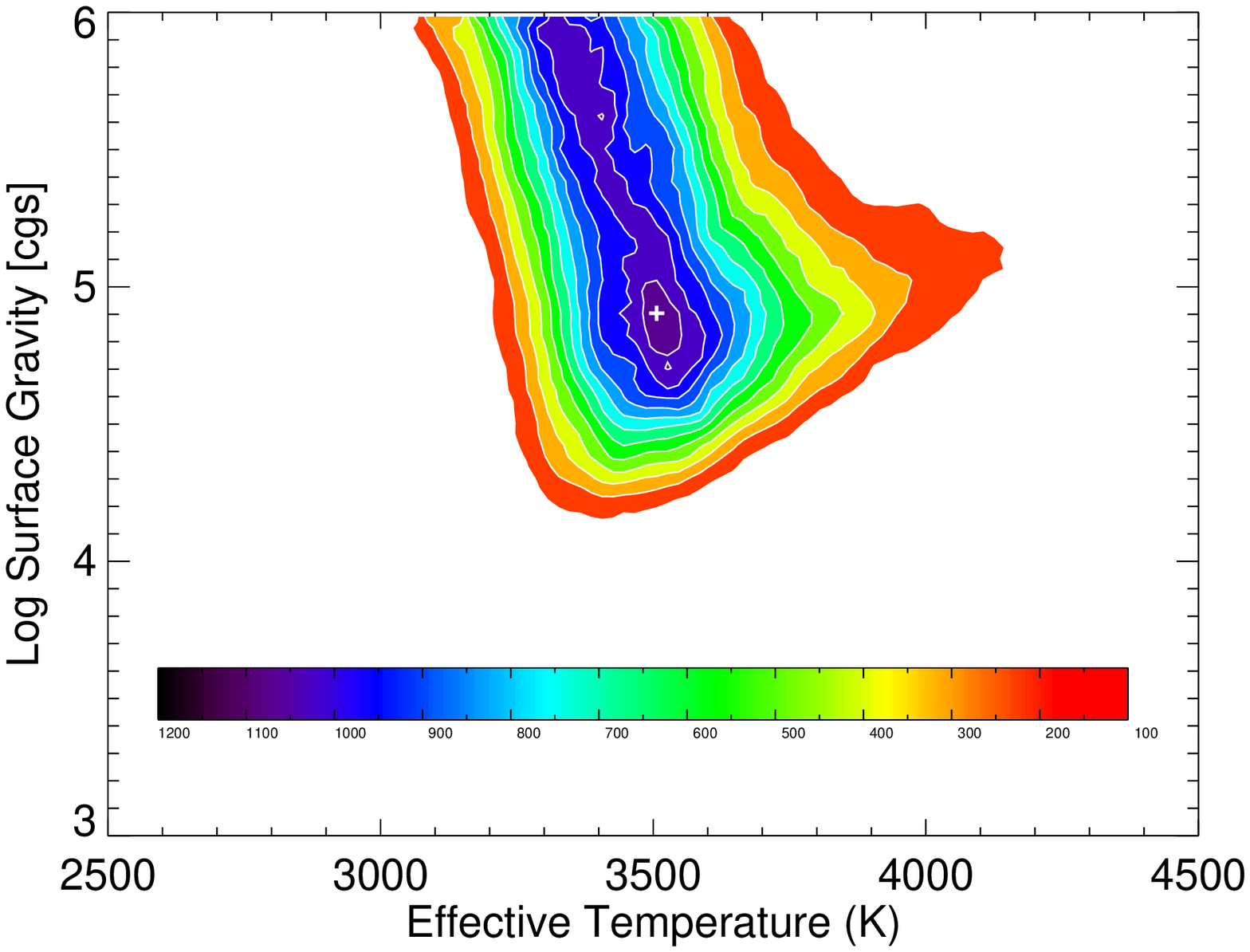}

  \end{center}

  \caption[]{ \label{hd129333b_contour} Density contours of Markov Chain Monte Carlo results for HD~129333 B in effective temperature (x-axis) and surface gravity (y-axis).  
}

\end{figure}

The parameter distribution provided by the MCMC routine are more informative about the potential of this analysis. Figure~\ref{hd129333b_contour} shows density contours of the accepted parameters from 9$\times$10$^5$ links in the MCMC. The highest contours in effective temperature are narrow and symmetric, providing a reliable (but large) 1-$\sigma$ uncertainty of 211~K. Although this is perhaps unpleasantly large, it is much smaller than the initial range of possible effective temperatures, 1400 to 4500~K, and did not require assumptions about the properties of the companion or the primary. The histogram for surface gravity, on the other hand, allows a large range of values, from $\sim$4.1~to 6.0~dex, the upper boundary of the model grid. The success of low-resolution near-infrared spectra at producing reliable temperature results, but not fairing so well in gravity is also shown by using simulated P1640 spectra created using templates from the IRTF spectral library (Rice et al., in prep.). Thus suggests that constraining the age of the system via detailed studies of the primary star will be necessary in order to reliably characterize directly-detected low-mass companions.

%%%%%%%%%%%%%%%%%%%%%%%%%%%%%%%%%%%%%%%%%%%%%%%%%%%%%%%%%%%%%%%

\section{Null Results}\label{null_results}

In addition to the five stars discussed above, we also observed 14 other stars that were part of the FEPS sample. These stars were observed in order to detect any additional companions than those previously known. We detected the known companion to HD 108944 at the edge of our field of view with only a part of the PSF was visible; we could not extract a spectra or measure accurate astrometry. For the rest of the stars, we did not detect any candidates in the our 3\farcs84 field of view. The companions of the known binaries are outside of our field of view.  Table \ref{unresolved} lists the HD number of the stars, the WDS number (indicating a known binary) and the Besselian date of the observations. In several cases, the stars were observed on multiple dates and we have listed all the dates.  After data reduction with speckle suppression software \citep{crepp2011}, all the stars had contrast curves approximately  the same as that shown in \citet{crepp2011}.

%%%%%%%%%%%%%%%%%%%%%%%%%%%%%%%%%%%%%%%%%%%%%%%%%%%%%%%%%%%%%%%

\section{Summary}\label{summary}

We studied five binaries with late type companions. For each companion we present the astrometry from several epochs as well as an estimate of the spectral type derived from matching measured spectra to template spectra.  Most of the binaries have periods of the order of centuries, but HD 129333 already has a computed orbit and additional astrometry is needed to refine this orbit. This is particularly important, as the orbit is a combined RV and visual orbit, allowing computation of the individual masses of the stars. Additional observations will improve the mass estimates. HD 135363 also appears to have a period on the order of several decades and additional observations will allow an eventual orbit computation. These results increase our understanding of these particular systems, and also help to validate the P1640 data collection and reduction processes. We have also gained valuable experience in estimating the physical parameters of low mass companions using synthetic spectra calculated using atmosphere models. While potentially powerful, this tool is subject to systematic uncertainties both from the observed spectra (i.e., poor telluric correction or unreliable points at the ends of the spectrum) as well as from the models and intrinsic sensitivity of the spectra to the physical parameters (i.e,. lack of precision in best-fit surface gravity at low spectral resolution).
  
In the future, we will collect multiple calibration observations in order to further reduce the spectral errors.  In addition, during there have been advances in data reduction pipeline which promise to further reduce the noise of the extracted data. While it is being developed for the Phase II data, we will apply the new pipeline to the Phase I data.  For Phase II data, improvements in the instrument have improved the optical throughput of the instrument, additionally decreasing the effect of noise.

%%%%%%%%%%%%%%%%%%%%%%%%%%%%%%%%%%%%%%%%%%%%%%%%%%%%%%%%%%%%%%%

\section{Acknowledgements}

In addition, we thank A.\@~Kraus and N.\@~Madhusudhan for useful discussions. The paper is based on observations obtained at the Hale Telescope, Palomar Observatory as part of a continuing collaboration between the California Institute of Technology, NASA/JPL, and Cornell University. We thank the staff of the Palomar Observatory for their invaluable assistance in collecting these data. 
A portion of the research in this paper was carried out at the Jet Propulsion Laboratory, California Institute of Technology, under a contract with the National Aeronautics and Space Administration (NASA) and was funded through the NASA ROSES Origins of Solar Systems Grant NMO710830/102190.   This research was also supported in part by the American Astronomical Society's Small Research Grant Program. In additsion, part of this work was performed under a contract with the California Institute of Technology funded by NASA through the Sagan Fellowship Program. Project 1640 is funded by National Science Foundation grants AST-0520822, AST-0804417, and AST-0908484. The members of the Project 1640 team are also grateful for support from the Cordelia Corporation, Hilary and Ethel Lipsitz, the Vincent Astor Fund, Judy Vale, Andrew Goodwin, and an anonymous donor.  This research made use of the Washington Double Star Catalog maintained at the U.S. Naval Observatory, the SIMBAD database, operated by the CDS in Strasbourg, France and NASA's Astrophysics Data System.  
 
%{\it Facilities:} \facility{Hale (Project 1640, PALAO)}, \facility{Keck:II (NIRSPEC2)} 

%%%%%%%%%%%%%%%%%%%%%%%%%%%%%%%%%%%%%%%%%%%%%%%%%%%%%%%%%%%%%%%

% References

% Astrometry
\begin{deluxetable}{lrcccl}
\tablewidth{0pt}
\tablecaption{Measured Astrometry\label{astrometry}}
\tablehead{\colhead{WDS} & \colhead{HD} & \colhead{Epoch} & \colhead{$\theta$ ($^{\circ}$)} &\colhead{$\rho$ (\arcsec)} &\colhead{Instrument} }
\startdata
09035+3750 & 77407 & 2002.0850 &   353.364 $\pm$ 0.043 & 1.6589 $\pm$ 0.0036 & PHARO\\
           &       & 2003.0351 &   353.796 $\pm$ 0.114 & 1.6685 $\pm$ 0.0032 & PHARO\\
           &       & 2003.3547 &   353.890 $\pm$ 0.030 & 1.6723 $\pm$ 0.0029 & PHARO\\
           &       & 2009.2083 &   355.843 $\pm$ 0.159 & 1.6875 $\pm$ 0.0071 & P1640\\
10368+4743 & 91782 & 2002.1669 &    33.572 $\pm$ 0.455 & 1.0022 $\pm$ 0.0084   & PHARO \\
           &       & 2003.0294 &    33.551 $\pm$ 0.265 & 1.0189 $\pm$ 0.0028   & PHARO \\
           &       & 2003.3544 &    33.833 $\pm$ 0.097 & 1.0250 $\pm$ 0.0021   & PHARO \\
           &       & 2004.0969 &    32.451 $\pm$ 0.107 & 1.0246 $\pm$ 0.0019   & PHARO \\
           &       & 2004.4280 &    32.532 $\pm$ 0.104 & 1.0322 $\pm$ 0.0007   & NIRC2\\
           &       & 2004.4305 &    32.639 $\pm$ 0.106 & 1.0243 $\pm$ 0.0010   & NIRC2\\
           &       & 2009.2056 &    30.5   $\pm$ 2.1   & 1.01   $\pm$ 0.08     & P1640\\
12547+2206 &112196 & 2002.0876 &    55.518 $\pm$ 0.094 & 1.5008 $\pm$ 0.0013   & PHARO \\
           &       & 2003.0350 &    54.839 $\pm$ 0.095 & 1.5114 $\pm$ 0.0012   & PHARO \\
           &       & 2003.3572 &    54.710 $\pm$ 0.086 & 1.5157 $\pm$ 0.0016   & PHARO \\
           &       & 2004.0972 &    54.351 $\pm$ 0.085 & 1.5241 $\pm$ 0.0026   & PHARO \\
           &       & 2004.4851 &    54.257 $\pm$ 0.097 & 1.5306 $\pm$ 0.0009   & PHARO \\
           &       & 2009.2086 &    50.62  $\pm$ 0.26  & 1.620  $\pm$ 0.008    & P1640\\
14390+6417 &129333 & 2003.0296 &   172.766 $\pm$ 0.114 &  0.7172 $\pm$ 0.0093  & PHARO \\
           &       & 2003.0323 &   172.783 $\pm$ 0.102 &  0.7286 $\pm$ 0.0019  & PHARO \\
           &       & 2003.3629 &   173.010 $\pm$ 0.318 &  0.7361 $\pm$ 0.0025  & PHARO \\
           &       & 2004.0974 &   172.522 $\pm$ 0.178 &  0.7415 $\pm$ 0.0032  & PHARO \\
           &       & 2009.2007 &   170.00  $\pm$ 0.55  &  0.764 $\pm$ 0.009    & P1640\\
15079+7612 &135363 & 2002.0877 &   121.352 $\pm$ 0.463 &  0.2506 $\pm$ 0.0027  & PHARO \\
           &       & 2003.0378 &   123.588 $\pm$ 0.354 &  0.2564 $\pm$ 0.0024  & PHARO \\
           &       & 2004.4882 &   127.384 $\pm$ 0.124 &  0.2846 $\pm$ 0.0007  & PHARO \\
           &       & 2009.2062 &    133.14 $\pm$ 0.83  &  0.363 $\pm$ 0.013    & P1640
\enddata
\end{deluxetable}

% Orbital residuals.
\begin{deluxetable}{lcc}
\tablewidth{0pt}
\tablecaption{The  $O-C$ Residuals for HD~129333 \label{residuals}}
\tablehead{
  \colhead{UT}  & \colhead{$\Delta$$\theta$ ($^\circ$)} & \colhead{$\Delta$$\rho$ (\arcsec)}
}
\startdata
2003.0296  &  -0.11 &  0.027\\
2003.0323  &  -0.11 &  0.027\\
2003.3629  &   0.18 &  0.039\\
2004.0974  &  -0.21 &  0.034\\
2009.2007  &  -2.09 &  0.015\\
\enddata
\end{deluxetable}

%%%%Fit results

\begin{deluxetable}{lclr}

\tablewidth{0pt}

\tablecaption{Spectral Results\label{results}}

\tablehead{

  \colhead{Name}  & \colhead{Sp. Type} & \colhead{T$_{eff}$ (K)} &  \colhead{Log(g) [cgs]} 

}

\startdata

HD~77407 B   &  M3V--M6V   &  3322$\pm$141 & 4.08$\pm$0.76   \\   

HD~91782 B   &  M9V        &  3246$\pm$63  & 3.02$\pm$0.07 [min] \\

HD~112196 B  &  M2V--M3V   &  3914$\pm$320 & 5.30$\pm$0.44    \\

HD~129333 B  &  M2V--M3.5V &  3506$\pm$211 & 4.90$\pm$0.46  \\

HD~135363 Ab  &  M2V--M4V   &  3667$\pm$281  & 5.90$\pm$0.21  [max]\\

\enddata

\end{deluxetable}

%%%%Fit results
\begin{deluxetable}{rcl}
\tablewidth{0pt}
\tablecaption{ Stars with No Detected Companions\label{unresolved}}
\tablehead{
  \colhead{HD}  & \colhead{WDS} & \colhead{Date (UT)} 
}
\startdata 
70516  & 08243+4457  &  2009.1973  \\
90905  & 10297+0129  &  2009.2056  \\
92855  & 10440+4612  &  2009.2058 \\
95188  &     ...     &  2009.2058\\
104860 & 12046+6620  &  2009.2087 \\
107146 &     ...     &  2009.2005\\
108944 & 12310+3125  &  2009.2060\\
134319 & 15058+6403  &  2009.2008\\
       &             &  2009.2035\\
       &             &  2009.2062\\
145229 &     ...     &  2009.4929\\
151798 & 16501-1223  &  2009.4903\\  
152555 &     ...     &  2009.2051\\
       &             &  2009.4875\\
172649 & 18397+3800  &  2009.4904\\
       &      ...    &  2010.5145\\
203030 & 21190+2614  &  2008.8083\\
204277 &     ...     &  2009.4881\\
       &             &  2009.4907\\
\enddata
\end{deluxetable}

\end{document}